\documentclass[twocolumn,showpacs,prl,amsmath,amssymb]{revtex4-1}
\usepackage{graphicx}% Include figure files
\usepackage{dcolumn}% Align table columns on decimal point
\usepackage{bm}% bold math

\begin{document}

\preprint{APS/123-QED}

\title{Stable Nanotubular Crystal of Silicon: A Predicted Allotrope With Direct Band Gap}% Force line breaks with \\

\author{Chi-Pui Tang, Jie Cao, and Shi-Jie Xiong}\email{sjxiong@nju.edu.cn}

\affiliation{National Laboratory of Solid State Microstructures and
Department of Physics, Nanjing University, Nanjing 210093, China}

\date{\today}% It is always \today, today,

\begin{abstract}
On basis of the first principle calculation we show that a crystalline structure of silicon, as a novel allotrope with nanotubular holes along two perpendicular directions, is stable. The calculations on geometrical and electronic properties reveal that this allotrope possesses a direct band gap wider by 0.5eV than the indirect one of silicon with diamond structure. The crystal belongs to I41/AMD space group, showing anisotropic optical properties and Young modulus. The bulk modulus is 64.4GPa and the density is 1.9g/cm$^{3}$, lower than that of the diamond silicon due to the presence of nanotubular holes. It is hopeful that the allotrope may widely expand applications of silicon in many fields due to its direct band gap and specific nanotubular structure.
\end{abstract}

\maketitle

 Crystal silicon (c-Si) of diamond structure plays the most important role in world semiconductor industry owing to its stability, purity, and ease of processing \cite{a1}. The shortage is its indirect band gap that restricts the applications in electro-optical devices. There are other allotropes of silicon with different structures and properties. Porous silicon (p-Si) is a form with nanoporous holes in its microstructure discovered by Uhlir and Uhlir in 1956. In 1990 Canham \cite{1} found the photoluminescence in p-Si at room temperature that attracted attention of scientists. Amorphous silicon (a-Si) is another allotropic form with non-crystalline structure that has been used as a solar cell material for its cheaper cost than c-Si. Recently, nanocrystalline silicon (nc-Si) \cite{a2}, an allotropic form with paracrystalline structure which has small grains of c-Si within the amorphous phase, has caused some interest in its use in tandem solar cell. However, all these allotropes, except the c-Si, have disordered structures which cause poor carrier mobility and low quantum efficiency.

From the first principle calculations, in this Letter we predict that the nanotubular crystalline silicon (ntc-Si), a new allotropic crystal, is stable and has a direct band gap wider than the indirect gap of c-Si by 0.5eV. Its geometry belongs to I41/AMD space group. The bulk modulus is 64.4GPa and the density is 1.9g/cm$^{3}$, both are lower than those of c-Si. Remarkably, ntc-Si has nanotubular holes along two perpendicular directions with section area 28.8 \AA $^{2}$ penetrating the whole crystal. The direct band gap and unusual nanotubular structures of this novel Si crystal provide a strong prospect of applications in electro-optical devices, new-style solar cells, photocatalyst, molecular sieves, and aerospace materials.

\begin{figure}[pt]
\centering
\includegraphics[width=7.5cm, keepaspectratio]{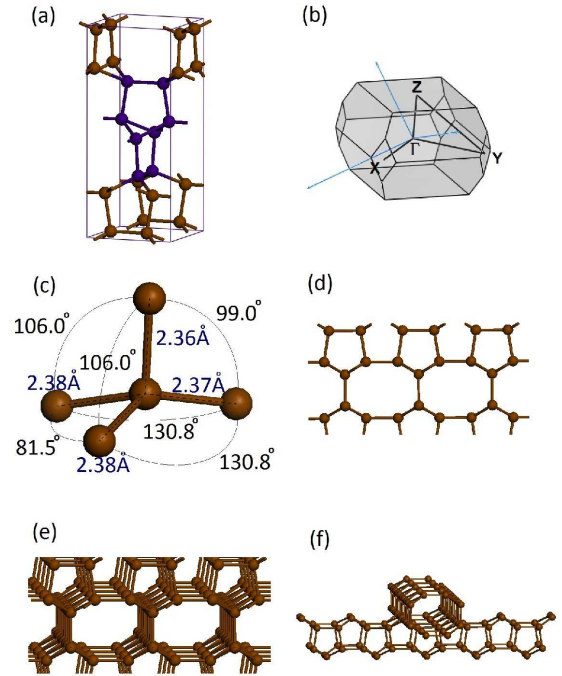}
\caption{(Color online) The crystalline structure of ntc-Si. (a): The body-centered unit cell of I41/AMD space group for ntc-Si. The blue balls indicate the central part which forms a primitive cell. (b): The Brillouin zone. (c): The bond angles and bond lengths of an atom with its 4 coordinates. (d): Pattern viewed along the [110] direction. (e): A part of ntc-Si crystal viewed along a direction slightly tilted from [110]. (f) Two cris-crossed octagonal nanotubes along the [110] and [1$\bar{1}$0] directions isolated from other atoms.}
\end{figure}

The calculations are perform on basis of the density functional theory (DFT) \cite{2,3} with the Cambridge sequential total energy package (CASTEP) \cite{a} using the ultrasoft pseudopotential method \cite{Vanderbilt}. To specify the effect of different functionals, we use two approximations: the local density approximation (LDA) developed by Ceperley and Alder \cite{4} and parameterized by Perdew and Zunger \cite{5} (CA-PZ), and the generalized gradient approximation (GGA) developed by Perdew {\it et al.} (PBE) \cite{6}. The kinetic energy cutoff of plane waves is 400eV. The Monkhorst-Pack scheme is used for the choice of $k$ points in the Brillouin zone \cite{7} and $12\times 12 \times 12$ $k$ points are taken for calculations of the band structure and properties. To reduce the systematic underestimation of the energy gap in DFT calculations \cite{aa3}, the band structure is also calculated by using B3LYP (Becke, 3-parameter, Lee-Yang-Parr) hybrid functional \cite{becke} and screened exchange functional (sX-LDA) \cite{sx-LDA}. The geometry optimization is performed using Broyden-Fletcher-Goldfarb-Shanno (BFGS) minimizer \cite{9} with the convergence tolerances set as 5.0$\times 10^{-6}$ eV/atom for energy, 0.01eV/{\AA} for maximum force, 0.02GPa for maximum stress, and 5.0$\times 10^{-4}${\AA} for maximum displacement. The ultrasoft pseudopotential is replaced by norm-conserving pseudopotential \cite{rappe} in calculations using B3LYP and sX-LDA functionals and in phonon calculations.

Initially we consider the structure of ntc-Si shown in Fig. 1. It is a body-centered structure and belongs to space group I41/AMD. Figure 1(a) displays the body-centered unit cell. In the central part of Fig. 1(a) indicated with blue balls, corresponding to a primitive cell, there are 8 atoms forming 2 pentagons with a common dehiscent head. The Brillouin zone is shown in Fig. 1(b). This structure ensures that each silicon atom has four coordinates as shown in Fig. 1(c). In the geometry optimization on this structure the convergence is rapidly reached, implying its stability. The optimized lattice constants of the unit cell are $a=b=5.45${\AA}, $c=13.00${\AA} from the GGA calculation and $a=b=5.37${\AA}, $c=12.7${\AA} from the LDA calculation. Through [110] direction we observe rows of octagons, separated with rows of pentagons [Fig. 1(d)]. This means that there are rows of nanotubes penetrating the whole crystal along the [110] direction as shown in Fig. 1(e). The octagonal section area of a nanotube is 28.82{\AA}. There are also rows of nanotubes with smaller octagonal sections in different planes. We can get the same projection pattern along the $[1\bar{1}0]$ direction. Note that the nanotubes along two directions are perpendicularly cris-crossing with a shifting along the $c$ direction, as indicated by Fig. 1(f). Such a hollow structure is reminiscent of the pores in p-Si, but ntc-Si is wholly crystalline and has much longer nanotubes with smaller sections.

Although all atoms are 4-coordinated, the short-range structure in ntc-Si is slightly different from that of the ordinary c-Si. As shown in Fig. 1(c), the bond angles and bond lengths for 4 bonds of an atom in ntc-Si are ${\angle}AOD=99.0^{\circ}$, ${\angle}AOC= {\angle}AOB=106.0^{\circ}$, ${\angle}COD={\angle}DOB=130.7^{\circ}$, ${\angle}COB=81.5^{\circ}$, $AO=2.35${\AA}, $ CO=BO=2.37${\AA}, and $DO=2.36${\AA}, slightly different from the regular tetrahedron in c-Si with the same bond length and bond angle ($\approx 109.5^{\circ}$). With both the LDA and GGA methods, the obtained differences of bond lengths in ntc-Si are within the range of 1\%, but the differences of bond angles may be up to 40\%. This implies that the covalence bonds in Si are rigid and the great difference between ntc-Si and c-Si crystals mainly comes from the large differences in bond angles.
The X-ray diffraction (XRD) with wavelength 1.54056{\AA} (copper spectrum) is simulated for ntc-Si structure obtained from the GGA method and the result is shown in Fig. 2, for the purpose of comparison with future experiments.

\begin{figure}[h]
\centering
\includegraphics[width=7cm, keepaspectratio]{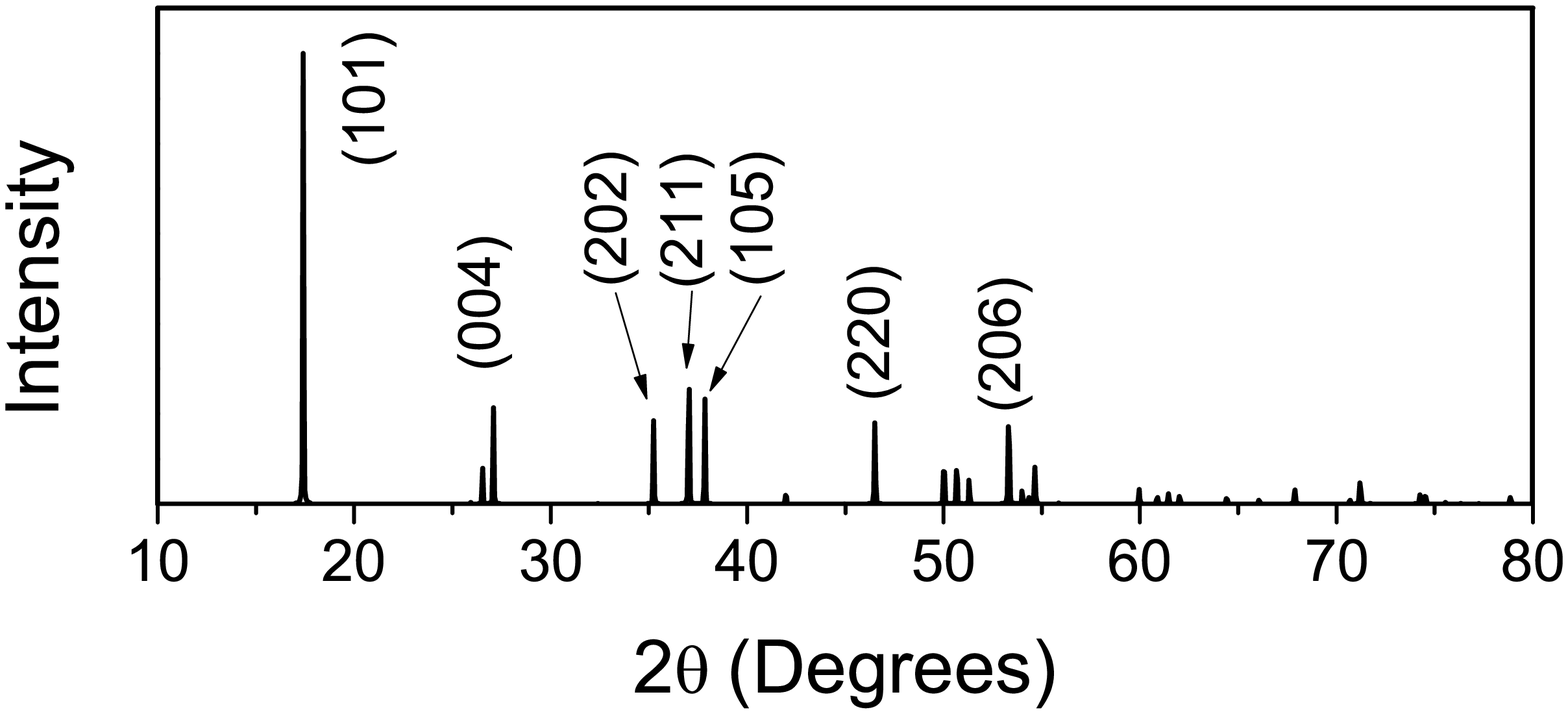}
\caption{ Simulated X-ray diffraction (XRD) with wavelength 1.54056{\AA} (copper spectrum) of ntc-Si structure obtained from the GGA method. }
\end{figure}

\begin{figure}[h]
\centering
\includegraphics[width=7cm, keepaspectratio]{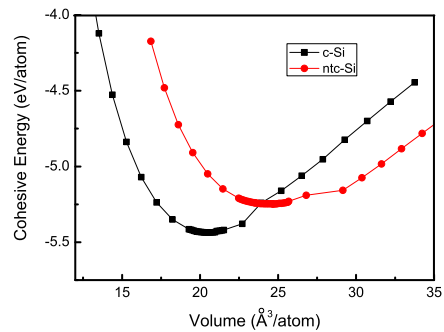}
\caption{(Color online) The cohesive energy per atom of Z-silicon and diamond-silicon dependence of volume per atom. }
\end{figure}

To estimate the stability of the structure, we calculate the cohesive energy per atom for both ntc-Si and c-Si by using the GGA method and the results are -5.24eV and -5.46eV, respectively. The results are -6.01eV and -5.81eV from the LDA calculation. Because the difference between them is only 0.2eV and c-Si is very stable, we can speculate that the stability of ntc-Si is also high in the same environment. To inspect the variation of the cohesive energy, we plot its dependence on volume per atom in Fig. 3 for both ntc-Si and c-Si. We can see that in both curves the energy minima occur in a similar way, the only difference is that for ntc-Si the minimum is slightly higher and appears at a larger volume per atom, suggesting that ntc-Si is a metastable phase with rather high stability.

To further examine the stability of ntc-Si, we calculate the phonon dispersion relation by using the method of linear response and norm-conserving pseudopotential and the results are displayed in Fig. 4. There are no anomalies, e.g., complex frequencies, in the phonon bands, strongly implying the mechanical stability of ntc-Si.

\begin{figure}[h]
\centering
\includegraphics[width=7cm, keepaspectratio]{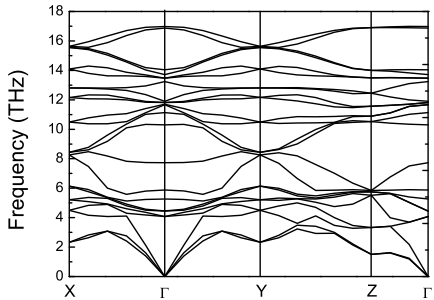}
\caption{Phonon bands structure of ntc-Si calculated with the linear response method and norm-conserving pseudopotential. }
\end{figure}

Then we investigate physical properties of ntc-Si and compare the results with c-Si. Several properties for them are listed in Table I. The density of ntc-Si is 1.93g/cm$^{3}$ from GGA calculation and 2.03g/cm$^{3}$ from LDA calculation, smaller than that of c-Si by about 0.3g/cm$^{3}$ due to the existence of hollow nanotubes. The bulk modulus of ntc-Si is 64.4GPa, also smaller than modulus 87.8GPa of c-Si. It is worth noting that the Young modulus in ntc-Si is anisotropic: it is 74.8GPa, 74.6GPa, and 116.4GPa in directions [100], [010], and [001], respectively, as obtained from the GGA calculation. The same calculation gives Young modulus 120.1GPa for c-Si in all 3 directions. In Table I we also present the results of modulus obtained from the LDA calculations. The anisotropy and smaller values of the Young modulus in ntc-Si are consistent with the expectations from the existence of the nanotubes along the [110] and [1$\bar{1}$0] directions.

\begin{table*}
\caption{Properties of ntc-Si: lattice constant $l$({\AA}), equilibrium density $\rho$(g/cm$^{3}$), bond length $d$({\AA}), cohesive energy $E_{coh}$(eV/atom), band gap $E_{g}$(eV), bulk modulus $B$(GPa), and Young modulus $Y$(GPa). }
\begin{ruledtabular}
\begin{tabular}{ccccccccc}
  Structure&Method&$l({\AA})$&$\rho$(g/cm$^{3}$)&$d({\AA})$&$E_{coh}$(eV/atom)&$E_{g}$(eV)&$B$(GPa)&$Y$(GPa)\\ \hline
 ntc-Si&GGA&$a=b$=5.45, $c$=12.99&1.93&2.36,2.37,2.38&-5.24&1.16&64.4&$a$=74.8,$b$=74.6, $c$=116.4 \\
 ntc-Si&LDA&$a=b$=5.37, $c$=12.74&2.03&2.31,2.33,2.33&-5.81&1.17&73.0&$a$=68.4, $b$=68.2, $c$=119.6 \\
 ntc-Si & H3LYP\footnote{The lattice constant of GGA is used by H3LYP calculation.}&$a=b$=5.45, $c$=12.99&--&--&--&1.25&--&-- \\
 c-Si&GGA&$a=b=c$=5.46&2.29&2.37&-5.4&0.60&87.8&$a=b=c$=120.1 \\
   c-Si&LDA&$a=b=c=$5.38&2.40&2.33&-6.0&0.44&96.9&$a=b=c$=126.7 \\
   c-Si&Exp \cite{Si2,si} & $a=b=c$=5.43&2.33&2.35&--&1.12&98.0&$a=b=c$=130.0 \\
  c-Si&H3LYP\footnotemark[1]&$a=b=c$ =5.46&--&--&--&0.71&--&--\\
\end{tabular}
\end{ruledtabular}
\end{table*}

\begin{figure}[h]
\centering
\includegraphics[width=7cm, keepaspectratio]{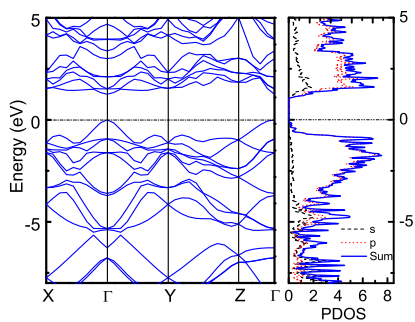}
\caption{(Color online) Electronic band structure and partial densities of states (PDOS) of ntc-Si calculated by using H3LYP functional and optimized structure. The direct band gap is 1.25eV. }
\end{figure}

One of the remarkable properties in ntc-Si is its direct band gap which is absent in c-Si. We calculate the band structure and the partial densities of states (PDOS) by using GGA, LDA, B3LYP, and sX-LDA methods. Only the results from the B3LYP calculation are displayed in Fig. 5. The bands structure evidently shows a direct band gap of 1.25eV in ntc-Si, while the same calculation gives an indirect gap 0.71eV for c-Si. The direct band gap in ntc-Si is wider than the indirect one in c-Si by 0.5eV. The results are almost the same from the sX-LDA calculation. The GGA and LDA calculations obtain values of band gap as 0.60eV and 0.44eV for c-Si, and 1.16eV and 1.17eV for ntc-Si, respectively. It is well-known that there is a systematic underestimation of the band gap in DFT calculations. We can use the difference of the calculated band gap from the experimental value in c-Si to correct the underestimation of the calculated band gap in ntc-Si. By this way we find that the band structure of ntc-Si is similar to that of crystalline GaAs in the direct gap and gap value. This may provide a method of producing semiconductors similar to GaAs but without introducing toxic As.

\begin{figure}[h]
\centering
\includegraphics[width=7cm, keepaspectratio]{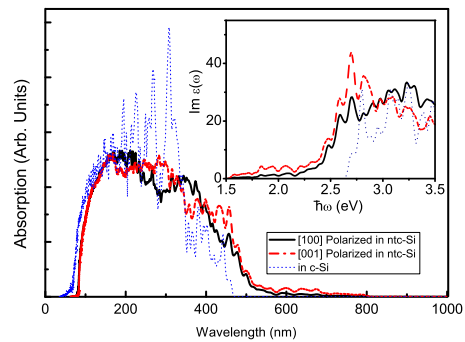}
\caption{(Color online) Optical absorption and the imaginary part of dielectric function of ntc-Si and c-Si calculated by using the H3LYP method. }
\end{figure}

We then calculate the optical properties for ntc-Si and c-Si by using the H3LYP method. The obtained absorption and imaginary part of dielectric function are exhibited in Fig. 6. Compared with the properties of c-Si in which the absorption is cut off at 480 nm and the imaginary dielectric function is cut off at 2.6 eV in the low energy region, the absorption and the imaginary dielectric function in ntc-Si are extended up to 800 nm and 1.5 eV, respectively. Considering that the band gap of ntc-Si is even larger than that of c-Si, we can conclude that the direct gap in ntc-Si much improves the optical activity in the low energy region. Such an improvement of the optical properties may extend applications of silicon crystals in electro-optical devices.

As a brief summary, we predict a novel stable crystal of silicon with hollow nanotubes penetrating through it. The stability is testified with calculations of the cohesive energy, geometry optimization, and phonon band structure. It is interesting that the structure has remarkable properties such as the regular penetrating pores, the direct band gap, the optical activities in a wide frequency range, and the anisotropy. As the structure of nanotubes can store or transport small molecules or atoms and the optical activities may be used in photo-chemical reactions, ntc-Si can serve as molecular sieves, electro-optical devices, new-style solar cells, photocatalyst, and aerospace materials.

\acknowledgments{ The authors thank W.J. Shi for helpful discussion. The work was supported by the State Key Programs
for Basic Research of China (Grant No. 2009CB929504), and by National Foundation of Natural Science in China of Grant Nos.  61076094 and 10874071.}

\end{document}